\documentclass[twocolumn,showpacs,secnumarabic,amssymb, amsmath, nofootinbib,tightenlines,nobibnotes, aps, prl]{revtex4}

\usepackage{graphicx}
\usepackage{dcolumn}
\usepackage{bm}

\begin{document}


\title{Direct Observation of a Temperature Dependent 0-$\pi$ Junction Transition in a Mesoscopic High Critical Temperature Grain Boundary Junction}

\author{G. Testa, A. Monaco, E. Esposito and E. Sarnelli }
\affiliation{Istituto di Cibernetica "E.Caianiello" del CNR, Via
Campi Flegrei 34, I-80078 Pozzuoli (Naples),Italy
}%

\author{D.-J. Kang, E.J. Tarte, S.H. Mennema and M.G. Blamire}
\affiliation{
Dept. of Materials Science, University of Cambridge,Pembroke Street, Cambridge,CB2 3QZ United Kingdom}%
 \altaffiliation[Also at ]{Nanoscience Centre, IRC in Nanotechnology, University of Cambridge, Cambridge CB3 0FF, United Kingdom}

\date{\today}

\begin{abstract}
We have observed, for the first time, a temperature dependent
0-$\pi$ junction crossover in a 45$^\circ$ symmetric [001] tilt
grain boundary junction. Experimental data, obtained by a double
phase-sensitive test based on a low inductance dc SQUID, show an
anomalous nonmonotonic temperature dependence of the Josephson
current, clear evidence of a critical current sign change at a temperature
T*. At the same temperature, the SQUID undergoes a transition from
a 0-SQUID to a $\pi$-SQUID. Results are in good agreement with
theoretical models taking into account mid-gap states in low
transparency junctions.
\end{abstract}

\maketitle

\begin{figure}
\includegraphics{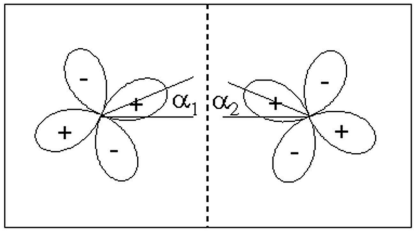}
\caption{\label{fig:epsart1} Schematic geometry of the grain
boundary interface. $\alpha_1$ and $\alpha_2$ are the angles
between the normal to the interface and the crystallographic axes
on the left and right sides, respectively.}
\end{figure}
\begin{figure}
\includegraphics{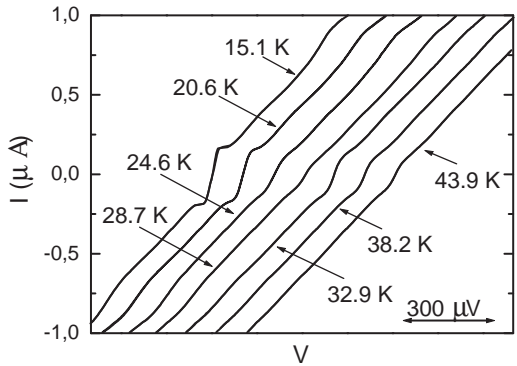}
\caption{\label{fig:epsart2} Current-Voltage characteristics of
the dc SQUID for different temperatures (from $T$ = 15.1 K to $T$
= 43.9 K). Curves have been shifted along the x-axis for sake of
clarity.}
\end{figure}
\begin{figure}
\includegraphics{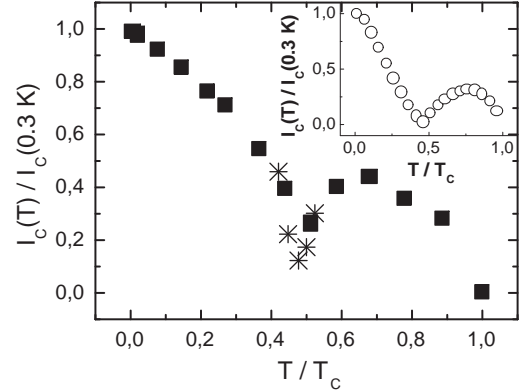}
\caption{\label{fig:epsart3} Normalized temperature dependence of
the Josephson current for the dc SQUID (squares, $2^{nd}$ set of
measurements and stars, $3^{rd}$ set of measurements). In the
inset we show the theoretical curve obtained by using the TK
formula with $\Delta_d(0) = 0.018$, $\kappa = 0.5$, $\lambda_0d_i
= 3$ and $\gamma = 0.2\Delta(0)$.}
\end{figure}
\begin{figure}
\includegraphics{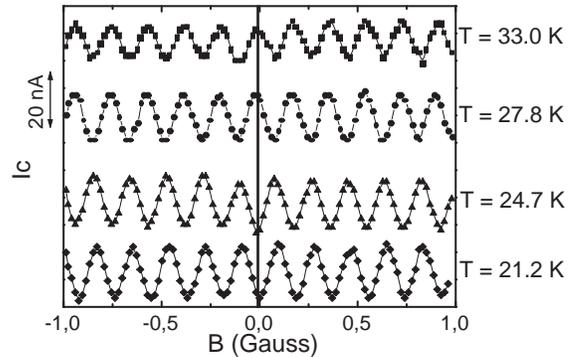}
\caption{\label{fig:epsart4} Magnetic field dependence of the
Josephson current as a function of the temperature. Curves have
been shifted along the y-axis for sake of clarity. The half flux
quantum shift between $T$ = 24.7K and $T$ = 27.8K clearly shows
the 0-$\pi$ transition of one of the two single junctions.}
\end{figure}

One of the most interesting debates in condensed matter physics is
the discussion about the symmetry of the superconducting wave
function in high critical temperature (HTS) cuprates
\cite{1,2,3,4}. In recent years, a number of experiments have
given a clear evidence of an unconventional d-wave order parameter
symmetry, characterized by a strongly anisotropic order parameter
with nodes along the (110) directions in $k$-space and a sign
change (corresponding to a $\pi$-phase shift of the
superconducting wave function) between orthogonal $k_x$ and $k_y$
directions. The most interesting results have been obtained by
phase-sensitive tests \cite{5,6,7,8}, mainly based on
superconducting loops containing one or more junctions, capable of
distinguishing between d-wave and asymmetric s-wave symmetries.
Another powerful phase-sensitive test is based on the property of
d-wave superconductors to form superconducting bound surface
states at the Fermi energy, the so-called mid-gap states (MGS)
\cite{9}. Such MGS, completely absent in conventional s-wave
superconductors, are generated by the combined effect of Andreev
reflections \cite{10} and the sign change of the d-wave order
parameter symmetry. MGS have been observed experimentally as a
zero bias conductance peak (ZBCP) in
superconductor/insulating/normal metal (SIN) and
superconductor/insulating/superconductor (SIS) junctions, and in
the impurity induced zero-energy density of states by scanning
tunneling microscopy measurements \cite{14,15,16,17,17b}.
\\One of the most striking consequences of MGS is the prediction of an
anomalous temperature dependence of the Josephson current
\cite{18,19,20}, strongly dependent on the barrier transparency
and on the misorientation angle $\alpha_i (i=1,2)$ between the
crystallographic axes of the two electrodes and the direction
normal to the interface. In particular, a transition from a 0 to
$\pi$-junction with decreasing temperature has been theoretically
predicted in 45$^\circ$ symmetric grain boundary junctions (GBJs)
with small transparency barriers. \\To date, no experimental
evidences of such a 0-$\pi$ transition has ever been reported in
the literature. In ref. \cite{21}, the authors reconstructed the
free energy of a mesoscopic 45$^\circ$ symmetric bicrystal GBJ as
a function of the phase $\varphi$, starting from current-phase
relation (CPR) measurements. They found, at low temperature, two
degenerate minima located at $\varphi = \pm \varphi_0$ (with $0 <
\varphi_0 < \pi$), in analogy with 45$^\circ$ asymmetric GBJs. The
authors related this double degenerate ground state to the
presence of an anomalously large second harmonic component of the
Josephson current, explained as the result of fluctuations of the
superconducting phase. Through CPR measurements, they also
extracted indirectly an anomalous temperature dependence of the
Josephson current, under the assumption that only two components
were present. \\In this Letter we show, for the first time, the
0-$\pi$ transition by direct transport measurements. Defining
$\theta$ as the angle between the quasiparticle trajectory and the
normal to the grain boundary interface, the order parameter sensed
by the quasiparticle incident on the barrier is $\Delta_i(\theta)
= \Delta_0 cos[2(\theta-\alpha_i)]$ (Fig. ~\ref{fig:epsart1}). In
45$^\circ$ symmetric junctions, MGS exist in both electrodes only
if quasiparticles probe different signs of the order parameter
before and after the reflection, i.e. only within limited
intervals of the angle $\theta$, $\theta/8 < |\theta| <
3\theta/8$. For other angles, the behaviour is similar to that of
s-wave superconductors. Because of the coupling between the two
electrodes, MGS form coherent states of the entire junction, the
Andreev Bound States \cite{18}. Andreev level energies are split
and shifted according to the phase difference $\varphi$ across the
contact and the barrier transparency $D$, where $0 < D < 1$.
Approximately, we can write $E_\pm = \pm |\Delta|\sqrt{D}sin
(\varphi/2)$. The lower energy level is then characterised by a
shift of $\pi$ in the ground state. At low temperatures, when only
the lower Andreev level is populated, the resonant contribution of
MGS, $I_{MGS}(\varphi)$, to the total Josephson current
$I_T(\varphi)$ dominates and becomes much larger than the
continuum state current, $I_{CONT}(\varphi)$. In fact,
$I_{MGS}(\varphi)$ is proportional to the square root of the
barrier transparency $\sqrt{D}$, as in any resonant coupling
phenomena [22], while $I_{CONT}(\varphi)$ is proportional to D. In
this case, the equilibrium phase difference of the whole junction
is $\varphi = \pi$. At higher temperatures, however, the two
Andreev levels give opposite currents and the MGS mediated current
is suppressed. The equilibrium phase difference is then $\varphi =
0$ as in conventional s-wave superconductors. Therefore, by
decreasing the temperature, the junction should undergo a
transition from an ordinary junction to a $\pi$ junction. The
critical current becomes zero at a temperature $T^j$ where
$I_{MGS}$ (negative) and $I_{CONT}$ (positive) balance each other.
\\In order to observe experimentally such a 0-$\pi$ transition,
we have designed a double phase-sensitive test. By using a low
inductance dc SQUID, we have investigated the 0-$\pi$ transition
by two independent assessments: a) the nonmonotonic temperature
dependence of the Josephson current, clear sign of a 0-$\pi$
transition at the temperature $T^*$ where the sum of the two
junction critical currents goes from positive to negative values;
b) a half flux quantum shift in the SQUID modulation at the
temperature $T^j$ where only one of the two junctions  (because of
the small asymmetry between the two grain boundaries, expected in
HTS GBJs) undergoes a transition from 0 to $\pi$. Of course, in
the absence of any 0-$\pi$ transitions, as in ref. \cite{21}, we
should find a different SQUID behaviour.
\\We have deposited 120 nm thick YBa$_2$Cu$_3$O$_{7-x}$ films on
45$^\circ$ symmetric [001] tilt SrTiO$_3$ bicrystal substrates by
pulsed laser deposition. The critical temperature $T_C$ ranged
between 89 and 91 K. A SrTiO$_3$ layer, 50 nm thick, and a gold
layer, 20-30 nm thick, have then been deposited to protect the
YBa$_2$Cu$_3$O$_{7-x}$ from the subsequent focused ion beam (FIB)
processing. The Au layer also prevents charging during the FIB
etch and inhibits degradation of the contact area during
photolithographic processes. Junctions with widths varying from 2
$\mu$m to 20 $\mu$m have been patterned by standard
photolithography and Ar ion milling. Three grain boundary
junctions, 0.8, 1 and 1.5 $\mu$m wide respectively and a dc SQUID
with two mesoscopic Josephson junctions, 0.3 $\mu$m wide, have
then been obtained by narrowing tracks using a FIB microscope with
a Ga source. Further details of the fabrication process will be
reported elsewhere. \\The reduction of the junction width limits
the number of defects at the grain boundary interface and
minimizes the influence of grain boundary faceting. In fact, it is
well known that 45$^\circ$ symmetric grain boundaries are
characterized by a large number of facets, with typical lengths of
10-100 nm; the angle $\theta$ formed by quasiparticles incident on
the barrier would then be averaged on many different facets making
the nonmonotonic dependence very difficult to observe.
\\Current-voltage (I-V) characteristics of both dc SQUID and
submicron GBJs have been measured as a function of the
temperature, from 1.2 K to $T_C$, by using a vacuum probe with a
heating element, covered by both a superconducting Pb cylinder and
a cryoperm shield. All the experiments have been made, with a very
low noise electronics, in a Helium cryostat shielded by one thick
aluminum and three mu-metal shields. In order to ensure that
trapped magnetic flux or external residual magnetic fields did not
influence the measurements of the Josephson current as a function
of the temperature we have repeated measurements three times
(after warming to room temperature), always obtaining the same
results. Samples have also been measured without any heaters, in
helium vapors, in order to rule out any possible noises or flux
trapping arising from the coupling between heater and signal
wires.\\At $T$ = 4.2 K, junctions show critical current densities
$J_C$ of the order of 3-5$\times$10$^3 A/cm^2$ and $I_CR_n$
products, with $R_n$ the asymptotic normal resistance, of about
300-500 $\mu V$; the dc SQUID has a lower critical current density
of about 3$\times$10$^2 A/cm^2$, a $R_n$ constant in the entire
temperature range and $I_CR_n = 40 \mu V$. The estimated Josephson
penetration length \cite{23} is of the order of 20 $\mu m$ and is
much larger than the widths of all the junctions.
\\Figure ~\ref{fig:epsart2} shows the I-V characteristics of the dc SQUID in the
temperature range 15 K - 44 K (I-V curves have been shifted along
the x-axis to allow a better comparison). The Josephson current
becomes almost zero at a temperature $T^*$ between 25 K and 28 K,
and increases for higher temperatures. This is the first time the
anomalous nonmonotonic temperature dependence of the Josephson
current, clear evidence of a current sign change, has been
observed by direct transport measurements. Even though the
critical currents are very small and the ratio between the
Josephson coupling energy $I_C\Phi_0/2\pi$ and the thermal energy
$K_BT$ is of the order of 0.01-1, the effect is unambiguously and
repeatably observable.
\\In Fig. ~\ref{fig:epsart3}
we report the dependence of the normalized Josephson current on
the temperature, $I_C(T)$, for the $2^{nd}$ set of measurements
(squares) and in a narrower temperature range around $T^*$,
$3^{rd}$ set of measurements (stars). Both data sets show a
minimum at the same temperature $T^*=0.5T_C$, and overlap
perfectly. It is worth noting that for the purpose of this
manuscript, every method to extract $I_C$ is equally valid. In
particular, we have extracted $I_C$ both by a derivative criterion
and by an extended RCSJ fit [21], taking into account the extra
contribution of the MGS mediated current, without finding
significant differences in the normalized curves.
\\The critical temperature of the dc SQUID is
about 56 K, much lower than the film $T_C$, as expected because of
the thermally activated phase slippage caused by the reduced
Josephson coupling energy at the grain boundary interface
\cite{24}. \\The observed temperature dependence of the Josephson
current can be well accounted for by theoretical models including
both the effect of the intrinsic phase of the pair potential and
the formation of localized states on the junction properties
\cite{18,19,20}. By assuming a rectangular barrier with thickness
$d_i$ and normalized height $k = K_F /\lambda_0$, where $K_F$ is
the Fermi momentum and $\lambda_0 = (2mU_0/\hbar^2)^{1/2}$, with
$U_0$ the Hartree potential at the interface, the Tanaka-Kashiwaya
(TK) formula gives the following $I(\varphi)$ \cite{19}:

\begin{eqnarray*}
R_nI(\varphi)=\frac{\pi R_nK_BT}{e} \left\{\sum_n
\int_{-\pi/2}^{\pi/2} F(\theta, i\omega_n, \varphi)\right.
\\ \times \left. sin(\varphi)
\sigma_n cos(\theta) d\theta \right\}
 \label{1}
\end{eqnarray*}

where $\sigma_n$ is the tunneling conductance for the injected
quasiparticle in the normal state, and $F(\theta,
i\omega_n,\varphi)$ is a function of $\Delta_{L(R)}$ (the absolute
value of the pair potential in the left (right) electrode), the
Matsubara frequency $\omega_n$, $\sigma_n$, $\kappa$, $\lambda_0$
and $d_i$.
\\In the inset of Fig. ~\ref{fig:epsart3}, we show the theoretical curve expected
for a 45$^\circ$° symmetric grain boundary junction with
$\lambda_0d_i = 3$ and $k$ = 0.5. In order to explain the
experimental saturation of the Josephson current at low
temperatures, we have considered the presence of defects at the
interface suppressing the MGS mediated current. A finite
quasiparticle lifetime parameter $\gamma = 0.2\Delta(0)$ has then
been introduced in the TK formula \cite{25} to describe the non
smooth interface. We can see a good qualitative agreement with
experimental curves. Any quantitative fits would not be physically
meaningful because of the presence of too many free parameters
($\kappa$, $\lambda_0d_i, \Delta_0, \gamma$).
\\It is worth noting that our Josephson
junctions, whose smallest width is 0.8 $\mu$m, a factor 2 or 3
larger than the junction width in the dc SQUID, showed a
conventional monotonic $I_C(T)$ dependence, with a slight upwards
curvature, like in ref. \cite{25}. This is an additional evidence
that the reduction of the junction width is mandatory to
investigate the effect of MGS on the Josephson current. It was
already demonstrated by CPR measurements on 45$^\circ$ symmetric
GBJs, which showed anomalous behaviours only for submicron widths
\cite{21,30}.
\\In order to gain a deeper insight into the physics of the junction transition
we have then investigated its influence on the properties of the
dc SQUID. In fact, by analyzing the SQUID dynamics, we can get
direct information about any phase changes in the two junctions.
Three different situations are possible: only one junction in the
dc SQUID undergoes a 0-$\varphi_0$ transition at $T^j$; both
junctions show the transition at the same $T^j=T^*$ value;
junctions show transitions at different values of $T^j (T^{j1}\neq
T^{j2})$.
\\In Fig. ~\ref{fig:epsart4}, we show the magnetic field dependence of the SQUID
critical current in a small temperature range across $T^*$ from 21
K to 33 K. Curves have been shifted along the y-axis for sake of
clarity. Diffraction patterns with a period of 0.2 G,
corresponding to an effective area of about 100 $\mu$$m^2$ and a
flux focusing $f = A_{eff}/A_{geom}$ of about 6, can be observed.
The $\beta$ value is of the order of $10^{-3}$ and so flux
trapping, self field effects and asymmetries are completely
negligible. The envelope of the SQUID modulations shows a Fraunhofer-like modulation pattern. It is worth noting that the residual field in our
cryostat is lower than 1 mG, corresponding to less than 0.01
$\Phi_0$.
\\The most interesting feature observed is the crossover from a
minimum to a maximum zero-field critical current, measured twice
with two different set-ups, at a temperature $T^j$ between $T$ =
24.7 K and $T$ = 27.8 K. This transition temperature is perfectly
consistent with the $T^*$ value extrapolated by I-V
characteristics. The half flux quantum shift is a strong evidence
that 1) only one of the two single junctions undergoes a phase
transition close to $T^*$; 2) the transition is not only a
0-$\varphi_0$ crossover but, with good accuracy, a transition from
a conventional state to a $"\pi"$ state, characterized by a shift
of $\pi$ in the energy ground state. In fact, (1) if both
junctions had the transition at the same $T^j=T^*$, the SQUID
would have remained in an unfrustrated configuration. Moreover
(2), if the state was $\varphi_0$, with $0 < \varphi_0 < \pi$, the
dynamics of the dc SQUID would have been completely different
\cite{27} and the SQUID modulations would not have shown an exact
$\Phi_0/2$ shift. Since $\varphi_0 = \pi$, we can assume that
harmonics larger than the first one give only a negligible
contribution to the Josephson current of our mesoscopic junctions.
Since the ratio $I_C^2/I_C^1$ between the second and the first
harmonics is proportional to $D$, a low-transmissivity is then
expected \cite{21}. By a qualitative fit of our $I_C(T)$
dependence with theoretical models, we can estimate a $D$ value of
the order of $10^{-2}$. This is only a little bit larger than the
average transmissivity estimated by the normal resistance value
$\rho_{ab}l/R_nA \approx 6\times10^{-3}$ \cite{28}, where
$\rho_{ab}$ (resistivity in the a-b plane) is about $10^{-4}
\Omega cm$ and $l$, the mean free path, is assumed equal to 10 nm.
It means that our grain boundary is uniform enough with only small
local changes. High transmissivity channels are then not expected
in our junctions.
\\In conclusion, by
a double phase-sensitive test, we have observed for the first time
the 0-$\pi$ transition theoretically expected in 45$^\circ$
symmetric GBJs. By using a low inductance dc SQUID we have
measured, by direct I-V characteristics, the anomalous
nonmonotonic temperature dependence of the Josephson current. Most
importantly, we have also observed a half flux quantum shift in
the magnetic field dependence of the SQUID critical current,
clearly confirming the 0-$\pi$ transition of one of the two single
junctions. Unlike other similar experiments on GBJs \cite{25}, the
barrier transparency was small enough and the interface
sufficiently clean to observe such an effect. It is worth noting
that our results support a strong d-wave component of the order
parameter. Indeed, highly symmetric SQUID oscillations with the
respect to zero magnetic field prove that any additional secondary
components (s or d$_{xy}$) give only a negligible contribution to
the predominant d-wave order parameter symmetry \cite{29}.

This work has been partially supported by the ESF Network
"Pi-shift", by MIUR under the project DG236RIC "NDA", by the TRN
"DeQUACS" and by EPSRC. We gratefully acknowledge discussions with
V.Shumeiko, A.Tagliacozzo and B.A.Davidson.


\begin{references}

\bibitem{1}R. Dagotto, Rev. Mod. Phys. {\bf 66}, 763 (1994).
\bibitem{2}D. J.
Scalapino, Phys. Rep. {\bf 250}, 330 (1995).
\bibitem{3}J. Annett, N. Goldenfeld and A.J. Leggett, "Physical Properties of High Temperature Superconductors", Ed. D. M. Ginsberg, (Singapore:
World Scientific 1996).
\bibitem{4}V.B. Geshkenbein, A.I. Larkin and A.Barone, Phys. Rev. B {\bf 36}, 235 (1987).
\bibitem{5}D. J. Van Harlingen, Rev.Mod. Phys. {\bf 67}, 515 (1995).
\bibitem{6}C. C. Tsuei and J. R. Kirtley, Rev.Mod. Phys. {\bf 72}, 969 (2000).
\bibitem{7}J.R. Kirtley et al., Nature {\bf 373}, 225(1995).
\bibitem{8}R.R. Schulz et al., Appl. Phys. Lett. {\bf 76}, 912 (2000).
\bibitem{9}C.R. Hu, Phys. Rev. Lett. {\bf 72}, 1526 (1994).
\bibitem{10}A.F. Andreev, Zh. Eksp. Teor. Fiz. {\bf 46}, 1823 (1964) (Engl. Transl. 1964 Sov.
Phys.-JETP {\bf 19} 1228).
\bibitem{14}M. Covington et al., Phys. Rev.Lett. {\bf 79}, 277 (1997).
\bibitem{15}I. Iguchi et al., Phys. Rev. B {\bf 62}, R6131 (2000).
\bibitem{16} H. Aubin et al., Phys. Rev. Lett. {\bf 89}, 177001 (2002).
\bibitem{17}L. Alff et al, Phys. Rev. B {\bf 58}, 11197 (1998).
\bibitem{17b}S. Kashiwaya et al, Phys. Rev. B {\bf 51}, 1350 (1995).
\bibitem{18}T. Lofwander, V.S. Shumeiko, and G. Wendin, Supercond. Sci.Technol. {\bf 14}, R53 (2001).
\bibitem{19}Y. Tanaka, and S. Kashiwaya, Phys. Rev. B {\bf 56}, 892 (1997).
\bibitem{20}Yu.S. Barash, Phys. Rev. B {\bf 61}, 678 (2000).
\bibitem{21}E. Il'Ichev et al., Phys. Rev. Lett. {\bf 86}, 5369 (2001).
\bibitem{22}G. Wendin and V.S.Shumeiko, Phys. Rev. B {\bf 53}, R6006 (1996)
\bibitem{23}A. Barone and G. Patern\`{o}, Physics and Applications of the
Josephson Effect (Wiley, New York, 1982).
\bibitem{24}R. Gross et al., Phys. Rev. Lett. {\bf 64}, 228 (1990).
\bibitem{25}H. Arie et al., Phys. Rev. B {\bf 62}, 11864 (2000).
\bibitem{30} E. Il'Ichev et al., Phys. Rev. Lett. {\bf 81}, 894 (1998).
\bibitem{27} T. Lindstrom et al., Phys. Rev. Lett. {\bf 90}, 117002 (2003).
\bibitem{28} G. Blonder, M. Tinkham, and T. Klapwijk, Phys. Rev. B {\bf
25}, 4515 (1982).
\bibitem{29} B. Chesca, Ann. Phys. (Leipzig) {\bf 8}, 511 (1999).
\end{references}
\end{document}